# Can Wearable Exoskeletons Reduce Gender and Disability Gaps in the Construction Industry?

Yana van der Meulen Rodgers, Rutgers University, Piscataway, NJ, USA

Xiangmin (Helen) Liu, Rutgers University, Piscataway, NJ, USA

Jingang Yi, Rutgers University, Piscataway, NJ, USA

Liang Zhang, New York University, New York, NY, USA

**Corresponding Author:** Yana Rodgers, Dept. Labor Studies & Employment Relations, Rutgers University, 94 Rockafeller Road, Piscataway, NJ 08854. Email yana.rodgers@rutgers.edu.

**Acknowledgments:** The authors acknowledge Douglas Kruse, Lisa Schur, Emma von Essen, Chunchu Zhu, and participants of the "Labor, Technology, and Carework" session at the 2025 Allied Social Science Associations Conference for their helpful suggestions. This research is supported by a grant from the National Science Foundation (NSF - ECCS-2222880).

**Abstract**
The share of construction trade jobs held by women and people with disabilities has remained stubbornly low in the face of chronic shortages of skilled labor. This study explores the potential of wearable assistive technologies to reduce these disparities. We use U.S. worker-level data to estimate employment and wage differences by gender and by mobility/strength impairments in construction and non-construction jobs. We also use occupational-level data to examine variations in workforce composition, physical skill requirements, and earnings across detailed construction occupations. Regression estimates indicate that being a woman and having strength and mobility impairments are associated with substantial employment and pay gaps in construction compared to non-construction jobs. Further analysis shows a high negative correlation between the representation of women and the ability levels required in those occupations. Finally, we discuss several wearable exoskeletons under development for people with upper-body and lower-body impairments, focusing on how these innovations could be integrated into construction jobs. These findings suggest that wearable exoskeletons that enhance manual dexterity, balance, and strength may improve the representation of women and people with disabilities in some of the higher-paying occupations in construction.

## Introduction

In recent decades, women have increased their representation in multiple non-traditional and historically male-dominated occupations such as automobile mechanics, police, firefighting, and airplane pilots (Zula 2014). Nevertheless, women's share of construction trade jobs has remained stubbornly low, rising only from 2.5% to 4.3% in the U.S. over the past two decades (BLS 2025a). People with disabilities also remain underrepresented in the construction industry: 8.1% of employed people with disabilities work in construction occupations, compared to 9.0% of people without disabilities (BLS 2025a). This underrepresentation is striking because construction is a major industry that offers relatively higher wages, stable employment, and pathways for skill advancement through apprenticeships and certifications, particularly for workers without college degrees.

The strenuous physical demands of construction work, including heavy lifting, repetitive overhead motion, and awkward postures, have long shaped recruitment and occupational norms. Indeed, an emphasis on above-average body strength and endurance continues to reinforce perceptions that women and people with disabilities are either physically unsuitable for or uninterested in such work (Fielden et al. 2000; Bailey et al. 2022). These assumptions contribute to their continued exclusion from construction occupations despite persistent labor shortages in the industry. Declining entry rates among younger workers and high attrition rates among incumbent workers have led to a chronic shortage of skilled workers since the 1980s, a trend projected to continue (Chini et al. 1999; Kim et al. 2020).

Assistive technologies offer a promising solution to address this labor shortage and improve the representation of women and people with disabilities in construction (Chao et al. 2024). Assistive technologies are products and systems that maintain or enhance an individual's functioning and independence to facilitate participation and improve well-being (Smith 2024). While most studies on workplace automation and the future of work have examined the displacement of human workers by machines, emerging research highlights the augmentation effect of new technologies, which can create new job tasks and enhance human capabilities, thereby giving labor a comparative advantage over capital (Acemoglu and Restrepo 2018, 2020). Assistive technologies inherently represent a form of human-technology collaboration and have the potential to redefine the boundaries of occupational segmentation based on manual strength by altering the physical requirements of work itself.

This study focuses on wearable exoskeletons, a class of assistive technologies designed to reduce strain and enhance strength through mechanical interaction with the body (De Looze et al. 2016). Engineering research on wearable exoskeletons has demonstrated benefits of biomechanical efficiency, strength enhancement, and task performance, primarily based on non-disabled participants in laboratory or pilot settings (Awolusi et al. 2018; Bär et al. 2021). Moreover, while social scientists have begun to examine the health and safety implications of wearable exoskeletons (Krzywdzinski et al. 2024), their role as technological innovations that expand employment access for underrepresented groups has received little attention.

We argue that long-standing assumptions on manual strength as a legitimate basis for occupational segmentation are increasingly incompatible with technological developments that can alter the physical demands of construction work. To advance understanding of the potential impact of exoskeletons in augmenting or equalizing physical capacity across workers, this study examines the empirical relationship between gender, physical ability, and employment access. Specifically, we first use worker-level data from the U.S. Survey of Income and Program Participation (SIPP) to estimate employment and wage differences by gender and mobility/strength impairments. We also compare these gender and mobility/strength gaps between construction and non-construction jobs. Next, we compile occupational-level data in the construction industry by merging the O*NET data from the Bureau of Labor Statistics (BLS) with public-use microdata from the American Community Survey (ACS). This enables us to examine variations in employment composition, physical skill requirements, and earnings across detailed construction occupations. Finally, we discuss several wearable exoskeletons under development for people with upper-body and lower-body impairments, focusing on how these innovations could be integrated into construction jobs.

## Background

Because of the large number of construction jobs in the U.S. economy, more women are employed in the construction trades than in other occupations, such as educational counsellors, social workers, librarians, pharmacists, and dental hygienists (BLS 2025a). Between 2021 and 2023 alone, the number of women working in construction trades increased by 15.7%, up to 363,351 workers (BLS 2025a). Compared to many female-dominated jobs, construction trades provide well-paid jobs that do not require a college degree. Yet the percentage of all construction occupations held by women has remained stubbornly low, staying below 5% for the past twenty years (Figure 1). Within construction, women's representation in production-oriented occupations is especially low (less than 20% of all jobs), while their representation is much higher in clerical and support positions (45.8%) (CPWR 2018). Production roles typically involve high levels of physical exertion, including lifting, bending, and repetitive motion, making them particularly relevant for wearable exoskeleton technologies. These roles also tend to have higher injury rates and more stringent physical skill requirements, which contribute to the exclusion of women and people with disabilities. In contrast, support roles generally involve less physical strain and are less likely to benefit directly from exoskeleton interventions.

**Insert Figure 1 Here**

The construction industry is notoriously dangerous, as it accounts for nearly one in five workplace fatalities; more than one-third of these deaths are caused by falls, slips, and trips (BLS 2021). Moreover, musculoskeletal disorders are prevalent among construction workers who frequently suffer from overexertion and repetitive motions. A substantial proportion of retirements from construction are not due to old age but rather work-related injuries and disabilities. Between 2007 and 2011, the construction industry lost nearly 2 million workers; in 2018, 80% of construction firms had trouble hiring skilled craft workers (Hearns 2019). Low motivation among young adults to enter and problems attracting and retaining non-traditional workers are key factors contributing to the persistent labor shortage (BLS 2020; Chini et al. 1999; Kim et al. 2020).

Occupational exoskeletons are designed to improve physical capability by supplying assistive torque and structural support at targeted joints, thereby shifting part of the mechanical demand from biological tissues to the device (Zhu et al. 2021). These mechanisms are consistent with evidence of sizable back, neck, and knee load reductions. For example, using back-assist exoskeletons was associated with a 24% reduction in hip extensor muscle activity and a 50% reduction in neck muscle strain, equivalent to lower perceived exertion during lifting and overhead tasks (Bosché et al. 2016). Wearable knee assistive devices decreased knee muscle activation by up to 39% and reduced knee-ground contact pressure by 15% while kneeling (Chen et al. 2021). These postures are common in construction tasks such as floor finishing, tiling, and rebar tying. Knee exoskeletons also help mitigate injuries and musculoskeletal disorders during awkward gaits, unexpected foot slips, and prolonged stance or kneeling (Zhu and Yi 2023; Sreenivasan et al. 2024).

While these studies have been tested primarily on workers who would otherwise be able to perform the tasks without exoskeletons, these effects are potentially beneficial for workers whose baseline strength or endurance is lower than that of incumbent construction workers. By lowering the peak force required to perform a task and delaying fatigue, wearable exoskeletons reduce the minimum strength threshold at which construction tasks can be executed within ergonomic risk limits. This relaxation of physical constraints enables workers in lower percentiles of the strength distribution to sustain kneeling, squatting, and overhead postures for longer durations, maintain precision with less co-contraction, and handle moderate loads more safely at standard repetition rates. This suggests that, if wearable exoskeletons were widely available and ready to accommodate diverse anthropometric characteristics and body masses, the disparities in achievable performance levels and injury risks among workers with varying physical capabilities would likely diminish. Under such conditions, the proportional benefits would be particularly high for women and workers with mobility or strength impairments, as exoskeletons would reduce the physical constraints that currently limit their participation in physically demanding jobs.

**Data and methodology**

The first part of this study uses microdata from SIPP to estimate how gender, physical mobility, and strength impediments are associated with employment and wage rates in construction versus non-construction jobs. Following the methods outlined by Kruse et al. (2024), we use the 2014 SIPP, the only wave that contains the Social Security Administration Supplement's detailed information about mobility and strength impairments. In particular, the data include nine indicators of difficulty with physical activities (climbing stairs, walking, standing, sitting, kneeling, reaching overhead, lifting, grasping, and pushing/pulling large objects), along with information on employment and earnings as well as demographic indicators such as education, race/ethnicity, and age. After restricting our sample to working-age individuals (18-64), we have 20,146 individuals, 6,554 of whom are not employed and 13,592 who are employed. The employed individuals are divided into two categories according to their industry: construction (1,001 observations) and non-construction (12,591).

We applied the SIPP data to a multivariable regression analysis of the determinants of employment and pay in the construction industry. These estimations are performed separately for construction and non-construction jobs. We use linear probability models to predict employment and a Heckman selection model to predict the natural log of hourly earnings. Independent variables include sets of dummy variables for gender, mobility/strength impairments, education, race/ethnicity, and age. We include control variables for nine physical impairments—difficulty climbing stairs, walking three blocks, standing or sitting for one hour, stooping/crouching/kneeling, reaching overhead, lifting/carrying ten pounds, grasping small objects, and pushing/pulling large objects—as well as four additional impairments related to vision, hearing, speech, and mental or cognitive functioning. Note that these categories do not map cleanly onto specific occupational requirements or job tasks within the construction industry, which makes it difficult to isolate the mechanisms through which physical impairments affect employment and earnings outcomes.

Our regression models do not include occupation fixed effects, which allows us to capture both within-occupation disparities and broader patterns of occupational sorting. This modelling choice partly reflects the relatively small sample size of construction workers in our dataset, which limits the feasibility of estimating reliable effects at the detailed occupational level. By retaining variation across occupations, we observe the overall employment and wage penalties associated with being a woman or having a disability in the construction industry—penalties that may reflect both unequal treatment within roles and restricted access to higher-paying, physically demanding jobs.

Consistent with Kruse et al. (2024), we adjusted hourly earnings at the top and bottom 1% of the earnings distribution by replacing extreme values with the values at those percentiles. The excluded variables that identify the Heckman equation are family size, number of children under age 18, other household income, and other household income squared. These models yield coefficients that indicate the percentage difference in employment and earnings associated with the key variables of interest, namely gender and mobility/strength impairments. These key indicators are binary variables in the separate regressions for construction and non-construction jobs. These regression results provide an upper-bound estimate of how wearable exoskeletons, designed to improve strength, dexterity, and range of motion, could draw more women into construction and increase their pay.

In the second part of the analysis, we merged the BLS O*NET data with public-use microdata from the ACS (2018 to 2022) to examine variations in employment composition, physical abilities, and earnings across jobs within construction. This approach expands upon the first part of our analysis in three ways. First, the ACS data is more recent than the SIPP, providing updated insights into employment and earnings. Second, the larger sample size in the ACS enables a more detailed examination of occupations within construction at the six-digit Standard Occupation Code (SOC) level, compared to the broader comparison of construction versus non-construction jobs in the SIPP data. Third, the O*NET data assesses 52 specific abilities across occupations (Peterson et al. 2001). The data on physical abilities encompass nine specific dimensions: stamina, dynamic flexibility, static flexibility, gross body coordination, gross body equilibrium, dynamic strength, explosive strength, static strength, and trunk strength. Each of these ability dimensions is measured between 0 and 7, with higher values representing higher levels of physical ability (Handel 2016). Finally, we calculate average annual earnings and the percentage of women in each occupation. Earnings are converted to 2022 constant dollars using the ACS earnings adjustment factor and the U.S. Consumer Price Index.

### Gender and mobility/strength gaps in construction versus non-construction jobs

Sample statistics from the SIPP data in Table 1 show that while women make up half of all non-construction jobs, they constitute just 8.3% of individuals employed in construction. By contrast, women are over-represented among individuals not working in the labor market, mainly due to their caregiving responsibilities. Also of note, of the nine types of mobility and strength impairments tracked in the SIPP data, seven impairments are associated with lower employment shares in construction compared to non-construction jobs. For example, 3.1% of individuals employed in non-construction jobs have difficulty climbing up ten flights of stairs, compared to just 1.6% of people in construction jobs. Only for standing and sitting impairments are the employment shares in construction higher, but the differences are close to zero. Consistent with other data reported by the Census, individuals with any impairments listed in Table 1 are more likely to be out of the labor force than employed (BLS 2025b). Also, construction workers have less education, are more likely to be Hispanic, are less likely to be young (ages 18-24) or mature (ages 55-64), and earn less than in non-construction jobs.

**Insert Table 1 Here**

Regression results in Table 2 in the first two columns show the probability change in construction employment (or non-construction employment) relative to the base of no employment, associated with a one-unit change in each indicator variable. Model 1 indicates that being a woman is associated with a substantial employment gap in construction (0.249 lower probability of employment in construction, or 24.9 percentage points, compared to men). None of the other explanatory variables in the construction regression produce a coefficient estimate of comparable magnitude. Model 2 suggests that the employment gap associated with being a woman is considerably smaller in non-construction jobs, at 7.9 percentage gaps.

**Insert Table 2 Here**

Table 2 further shows that individuals with four different types of mobility/strength impairments experience statistically significant employment gaps in construction compared to individuals without such impairments: difficulty walking, standing, stooping/crouching/kneeling, and pushing/pulling large objects. Three additional impairments are associated with employment gaps only in non-construction jobs: climbing stairs, reaching overhead, and lifting/carrying 10 pounds. Given that these physical activities are quite common in many construction jobs, it is possible that the relatively small sample size for construction jobs could explain the imprecise estimates in the construction employment regression. Other results in the employment regressions are as expected. Individuals with higher education are less likely to be employed in construction than those with a high school degree or less. Individuals who are Black non-Hispanic, Asian, or multiracial/other are relatively less likely to be employed in construction, and people above the age of 25 are more likely to be in construction compared to young adults ages 18 to 24.

Models 3 and 4 of Table 2 show earnings penalties or premiums associated with each indicator. In construction, being a woman is associated with a 0.356 log-point earnings penalty relative to men, substantially larger than the 0.231 log-point penalty observed in non-construction occupations. Among the mobility/strength impairments, only difficulty with stooping/crouching/kneeling is associated with a statistically significant earnings penalty in construction jobs. Interestingly, individuals working in construction jobs who have trouble walking earn a premium (0.442 log points), which could be explained as an occupational effect within construction: people who have difficulty walking may be more likely to have administrative jobs that pay more. In terms of race and ethnicity, Hispanic and other race individuals experience sizeable pay penalties compared to white non-Hispanic people, and these penalties are similar in magnitude to the penalty experienced by women. The remaining results are intuitive: in construction and non-construction jobs, individuals with higher education enjoy earnings premiums compared to people with high school or less, and those premiums rise with greater educational attainment. In addition, there is a positive earnings gradient that comes with age.

Table 3 reports results for ability requirements, pay, and women's representation across detailed occupations within the construction industry. Women in construction are most likely to work as hazardous materials removal workers, inspectors, painters, or paperhangers. Their representation among hazardous materials removal workers is particularly high (double that of the next category) and may be explained by the fact that some jobs within this category, such as asbestos and lead-based paint abatement, do not require the same physical strength as other construction-related jobs. Overall, occupations with higher female representation tend to pay less than those dominated by men, such as elevator installers, derrick and rotary drill operators, mining machine operators, and steel workers. A simple correlation analysis reveals that occupations with higher percentages of women workers have a lower income level (r=-0.413) and lower ability levels (r=-0.331). The implication is that technologies that enhance women's manual dexterity, balance, and physical strength can potentially improve their representation in some of the higher-paying occupations with more strenuous physical ability requirements.

<div align="center">**Insert Table 3 Here**</div>

### Examples of wearable exoskeletons

As the empirical results demonstrate, physical activity limitations can prevent people from getting or keeping jobs in the construction industry. They can also contribute to substantial earnings penalties. Recent advances in wearable exoskeleton technology show promise for narrowing these gaps and enhancing worker safety and health (Bär et al. 2021; Okpala et al. 2022). Below, we present three groups of wearable exoskeletons for construction work, categorized by their functional focus on (1) the back, (2) shoulders and arms, and (3) lower limbs. Although their mechanisms vary, all share a common principle: redistributing joint moments from vulnerable musculature to the device and to stronger proximal segments (Crea et al. 2021). This redistribution reduces peak torque demands, slows fatigue accumulation, and lowers the minimum strength threshold required to perform tasks at standard repetition rates. These effects are especially beneficial for workers with lower baseline strength, including many women and individuals with disabilities, for whom such physical demands often constitute a barrier to entry and retention in construction employment.

*Case 1: Back support exoskeletons*
Passive back-support exoskeletons are lightweight, battery-free devices designed to reduce strain on the lower back by transferring weight to the hips during lifting and bending tasks. Researchers have evaluated the HeroWear Apex in controlled trials, a commercially available passive back-support exoskeleton. Lamers et al. reported reductions in erector spinae (ES) muscle activity of 23-43% during leaning tasks and 14-16% during lifting tasks (Lamers et al. 2018), with subsequent evidence of significant reductions in muscle fatigue during leaning tasks (Lamers et al. 2020). Goršič et al. (2021) observed a 15% decrease in erector spinae electromyogram activity during object lifting and lowering, with participants reporting the exosuit as mildly to moderately helpful. These findings indicate that passive back-support exoskeletons can lower the minimum back-strength threshold for common tasks and extend workers' time to sustain forward-bent posture without exceeding ergonomic risk limits.

Active back-support exoskeletons, in contrast, are powered systems that deliver programmable assistive torque at the hip or lumbar region, sustaining support through long duty cycles and variable task phases. Sposito et al. (2024) conducted a multi-day field assessment of an active back-support exoskeleton (StreamEXO) with railway construction workers, who used the device for approximately 90 minutes on three non-consecutive workdays. Results showed a positive correlation between self-reported fatigue reduction and exoskeleton use during physically demanding movements. Moreover, qualitative data suggest that weight balance, body pressure, and thermal comfort influence user comfort and acceptance.

In a systematic review of over thirty studies on passive and active back-support exoskeletons, Kermavnar et al. (2021) reported that back support exoskeletons generally reduce back-muscle activity and improve endurance during lifting and static bending. However, performance tends to decline in tasks requiring greater agility. While most evaluations were conducted in laboratory settings with healthy young male participants, these findings provide support for the potential for back support exoskeletons to enhance the ability of women and workers with impairments to safely and effectively perform material-handling and forward-bending construction tasks such as floor layout, rebar tying, deck nailing, and box handling.

*Case 2: Shoulder and arm support exoskeletons*
Passive shoulder-support systems rely on gravity compensation and mechanical elements such as springs, elastic bands, cams, and Bowden cables to redistribute part of the shoulder moment to the torso or hips. This design reduces torque demands on the deltoid and scapular stabilizers during elevated-arm postures. Empirical studies show that passive systems consistently lower shoulder muscle activity (EMG) and perceived exertion in overhead drilling and assembly tasks, with the most potent effects observed during sustained ceiling-level work. For example, Bennett et al. (2023) examined the impact of the Hilti EXO-S on shoulder flexion (raising the arm to the front) and extension (pushing the elbow to the back of the body) and found a reduced 9-95% range of motion (ROM) in the shoulder. They also noted a slight reduction in the time required for tasks like pushing/emptying gondolas and installing/removing wooden blocks.

Likewise, Kim et al. (2018a, 2018b) reported that the EksoVest decreased shoulder muscle activity by up to 45%, particularly during overhead tasks, and reduced drilling task completion times by 20%. Although passive devices are lightweight, durable, and cost-effective, researchers have noted user discomfort from straps and pressure points, as well as potential kinematic mismatches with the complex scapulohumeral rhythm. These limitations can diminish perceived workload benefits, even when reductions in EMG activity are observed (Reyes et al., 2023).

Active shoulder exoskeletons incorporate motors or series-elastic actuators to modulate assistance dynamically across the arm-elevation cycle. Experimental evaluations indicate that active systems can adapt support to task phases and user preferences, improving flexibility for variable tools, angles, and loads while reducing shoulder stress and preserving movement quality (Reyes et al., 2023). However, their deployment on jobsites remains limited due to added weight, cost, and penalties related to mass, inertia, power consumption, and thermal or noise output (Schiebl et al., 2025). These challenges further constrain use in construction environments, where powered shoulder exoskeletons can interfere with personal protective equipment and tool handling in workplaces exposed to dust, moisture, and heat.

Evidence indicates that shoulder exoskeletons provide the greatest benefits for tasks that require holding the arms at or above shoulder height, such as drywall installation, overhead fastening, hanger and pipe runs, and long-reach painting. Benefits diminish when tasks require frequent changes in working planes or large reach arcs (de Vries et al., 2021; 2023; Bennett et al., 2023). In addition, fit, ROM, and comfort remain critical determinants of adoption, particularly for women and workers with limited shoulder endurance (Reyes et al., 2023).

**Case 3: Lower-limb exoskeleton**

Prototype knee-support exoskeletons for construction tasks are designed to mitigate the biomechanical demands of kneeling, squatting, and posture transitions common in floor finishing, tiling, and rebar tying. These activities impose high compressive and shear forces on the knee joint, contributing to musculoskeletal disorders and long-term degenerative conditions such as osteoarthritis (Sreenivasan et al. 2024). Laboratory and pilot studies demonstrate that knee exoskeletons can reduce knee extension/flexion muscle activity by up to 39% and knee-ground contact forces by approximately 15% during kneeling, indicating a substantial reduction in cumulative load on the knee complex during prolonged ground-level work (Chen et al. 2021). Such reductions are particularly relevant for workers with limited lower-limb strength or pre-existing joint conditions.

Beyond static kneeling support, Yi and collaborators have advanced research on dynamic stability and fall prevention, which are of critical concern in construction environments. For example, Zhu and Yi (2023) developed bilateral knee exoskeletons integrated with inertial sensing for real-time slip detection and adaptive torque control, thereby enabling rapid recovery from unexpected foot slips in wet or low-friction jobsites. Their experiments show that exoskeleton-enabled recovery strategies significantly shorten slip recovery time and reduce slip kinematics. In another study on neural balance strategies during quiet stance and kneeling under destabilizing conditions, Sreenivasan et al. (2024) demonstrated that knee exoskeleton assistance reduces center-of-pressure sway area by up to 62% in quiet stance and 39% in kneeling, highlighting the feasibility of exoskeleton-based interventions for fall risk mitigation. Such postural stability improvements enhance safety and reduce barriers for workers who may have balance limitations.

Although knee-focused exoskeletons are still experimental, their demonstrated potential points to a promising avenue for enabling workers with diverse physical capabilities to perform demanding tasks such as load carriage and repetitive lifting in sack handling, masonry, and scaffolding installation.

## Discussion

The results of this study reveal substantial employment and wage gaps in the construction industry for women and individuals with strength and mobility impairments. Compared to non-construction jobs, these disparities are more pronounced, suggesting that physical skill requirements continue to act as barriers to entry and advancement. The strong negative correlation between the representation of women and the physical ability levels required in construction occupations underscores the structural challenges that limit inclusivity in this sector. Wearable exoskeletons that enhance manual dexterity, balance, and strength may improve the representation of women and people with disabilities in some of the higher-paying occupations in construction.

These findings align with prior research documenting gendered and ableist barriers in the construction industry. Recruitment practices that emphasize above-average upper-body strength requirements discourage women and people with disabilities from applying for construction jobs, while safety and health hazards limit their retention (Fielden et al. 2000; Bailey et al. 2022). Our study builds on this literature by quantifying the employment and wage gaps and linking them to specific occupational skill requirements, thereby offering a more granular understanding of some of the mechanisms behind exclusion.

Similar patterns have been observed internationally; for example, Bolghanabadi et al. (2024) review occupational health and safety studies from Iran, Canada, and Australia, showing that musculoskeletal disorders and ill-fitting PPE disproportionately affect women in physically demanding sectors like construction. Likewise, Baghdadi (2024) highlights how infrastructure projects in developing countries often fail to accommodate the physical needs of women and disabled workers, with inadequate safety measures and strength-based job designs contributing to exclusion. These cross-national findings suggest that structural barriers to inclusion are not unique to the U.S. context but are embedded in the global construction industry.

The integration of wearable exoskeletons into construction work presents a promising avenue for reducing these disparities. However, for these technologies to fulfil their inclusive potential, they must be adaptable to a range of body types and movement patterns. Ensuring a proper fit for exoskeletons is critical for both safety and usability. Historically, work-related equipment, including personal protective gear, has been designed based on data from male military recruits or industrial workers from the mid-20th century. This design approach fails to accommodate the diverse body shapes and sizes in today's workforce, including women and people with disabilities (Søraa and Fosch-Villaronga 2020). Even with multiple size options, proper fit alone does not guarantee comfort or usability, especially for women. Women's bodies differ from men's in size and movement patterns, friction points, and areas of sensitivity. Common issues include bulky breastplates, poorly positioned chest pads, and incorrect proportions that fail to account for the average woman's torso length or hip width. These design flaws can result in awkward postures, increasing the risk of musculoskeletal strain, particularly in the shoulders and back.

In response to these challenges, recent research and commercial developments have focused on making exoskeletons more inclusive and adaptable for diverse users, particularly women (Gutierrez et al. 2024). Beyond issues of sizing and fit, workers may encounter difficulties with donning and doffing the devices, especially in dynamic and time-sensitive jobsite environments. Additionally, the cost of exoskeletons remains a significant barrier to widespread adoption, particularly in small and medium-sized construction firms where employers may be reluctant or unable to invest in such technologies without clear evidence of return on investment. Government agencies and innovation funders can help to support inclusive design and subsidize the development and deployment of these technologies, especially in sectors where market incentives alone may not prioritize equity (Clarke et al. 2009; Abdi et al. 2021).

The study has several limitations. Most notably, the regression analyses rely on 2014 survey data, which would not capture any potential gains made by women and people with disabilities in the past decade. In addition, our study does not establish a causal link between physical skill requirements and the underrepresentation of women and people with disabilities in the construction industry. While our regression results show strong correlations between physical impairments and employment and pay, the data do not allow us to isolate the direct effects of physical capacity from other factors such as harassment, discrimination, and workplace culture that are also linked to the low representation of women and people with disabilities in construction (Eisenberg 2018; Morello et al. 2018; Bailey et al. 2022). As such, the findings should be interpreted as indicative of patterns of exclusion rather than definitive evidence of causality. Another limitation is that our data include broad measures of physical difficulties, limiting our ability to assess how specific types of disabilities interact with occupational demands in the construction industry. Finally, our study does not include direct user feedback or field trials of wearable exoskeletons, which are essential for assessing real-world usability and impact.

Nonetheless, the observed associations provide a useful starting point for considering how wearable exoskeletons might mitigate some of the physical barriers that contribute to occupational sorting and wage disparities in the construction industry. Future research should incorporate more recent and longitudinal data to capture trends in workforce composition, device adoption, and user experiences among women and people with disabilities. While a growing number of studies have identified problems of exoskeleton fit for women, less is known about how exoskeletons can be adapted to people with disabilities. Drawing on data from the British and Dutch construction sectors, Clarke et al. (2009) show that a substantial proportion of long-term disabled workers in construction acquired their impairments because of work-related injuries. This finding underscores the need to broaden disability inclusion strategies in construction: wearable assistive technologies should not only be designed to help unemployed persons with disabilities gain employment, but also serve as enabling tools that support previously employed, now-impaired workers to remain in or return to employment after injury. Moreover, field-based studies that evaluate exoskeleton performance across diverse construction tasks and user groups would provide valuable insights into practical implementation. Future research could also expand on this study by exploring a broader range of assistive technologies beyond exosuits, such as ergonomic tools, adaptive personal protective equipment (PPE), and digital navigation aids, which may address other barriers to entry and retention in construction, particularly for workers with sensory or cognitive impairments. Finally, more granular data linking specific impairments to job tasks and workplace accommodations would enable deeper analysis of how assistive technologies can be tailored to diverse needs.

workers: a field-based pilot study. *Buildings* 13(3), 822. https://doi.org/10.3390/buildings13030822.

Bolghanabadi S, Haghighi A, Jahangiri M (2024) Insights into women's occupational health and safety: A decade in review of primary data studies. *Safety*, 10(2), 47. https://doi.org/10.3390/safety10020047

Bosché F, Abdel-Wahab M and Carozza L (2016) Towards a mixed reality system for construction trade training. *Journal of Computing in Civil Engineering* 30(2), 04015016. https://doi.org/10.1061/(ASCE)CP.1943-5487.0000479.

Bureau of Labor Statistics (2020) *National Census of Fatal Occupational Injuries in 2019*. Technical Report, Bureau of Labor Statistics, Washington, DC. Available at https://www.bls.gov/news.release/archives/cfoi_12162020.pdf (accessed 2 August 2025).

Bureau of Labor Statistics (2021) Construction deaths due to falls, slips, and trips increased 5.9 percent in 2021. *The Economics Daily.* Available at https://www.bls.gov/opub/ted/2023/construction-deaths-due-to-falls-slips-and-trips-increased-5-9-percent-in-2021.htm (accessed 2 August 2025).

Bureau of Labor Statistics (2025a) *Labor Force Statistics from the Current Population Survey*. Available at https://www.bls.gov/cps/tables.htm (accessed 4 August 2025).

Bureau of Labor Statistics (2025b) *Persons with a Disability: Labor Force Characteristics — 2023*. Available at https://www.bls.gov/news.release/archives/disabl_02222024.pdf (accessed 2 August 2025).

Chao GT, Deal C, Migliano EN (2024) Occupational exoskeletons: Supporting diversity and inclusion goals with technology. *Journal of Vocational Behavior* 153, 104016. https://doi.org/10.1016/j.jvb.2024.104016.

Chen S, Stevenson DT, Yu S, Mioskowska M, Yi J, Su H and Trkov M (2021) Wearable knee assistive devices for kneeling tasks in construction. *IEEE/ASME Transactions on Mechatronics* 26(4), 1989-1996. https://doi.org/10.1109/TMECH.2021.3081367.

Chini AR, Brown BH and Drummond EG (1999) Causes of the construction skilled labor shortage and proposed solutions. *ASC Proceedings of the 35th Annual Conference*. San Luis Obispo, CA: California Polytechnic State University, 187-196. Available at http://ascpro0.ascweb.org/archives/1999/chini99.htm (accessed 2 August 2025).
17

20Krzywdzinski M, Evers M, and Gerber C (2024) Control and flexibility: the use of wearable devices in capital-and labor-intensive work processes. *ILR Review* 77(4), 506-534. https://doi.org/10.1177/00197939241258206

Lamers EP, Yang AJ and Zelik KE (2018) Feasibility of a biomechanically-assistive garment to reduce low back loading during leaning and lifting. *IEEE Transactions on Biomedical Engineering* 65(8), 1674-1680. https://doi.org/10.1109/TBME.2017.2761455.

Lamers EP, Soltys JC, Scherpereel KL, Yang AJ and Zelik KE (2020) Low-profile elastic exosuit reduces back muscle fatigue. *Scientific Reports* 10(1), 15958. https://doi.org/10.1038/s41598-020-72531-4.

Morello A, Issa RR and Franz B (2018) Exploratory study of recruitment and retention of women in the construction industry. *Journal of Professional Issues in Engineering Education and Practice* 144(2), 04018001. https://doi.org/10.1061/(ASCE)EI.1943-5541.0000359.

Okpala I, Nnaji C, Ogunseiju O and Akanmu A (2022) Assessing the role of wearable robotics in the construction industry: potential safety benefits, opportunities, and implementation barriers. *Automation and Robotics in the Architecture, Engineering, and Construction Industry*, 165-180. https://doi.org/10.1007/978-3-030-77163-8_8.

Peterson NG, Mumford MD, Borman WC, Jeanneret PR, Fleishman EA, Levin KY, ... and Dye DM (2001) Understanding work using the Occupational Information Network (O* NET): Implications for practice and research. *Personnel Psychology* 54(2), 451-492. https://doi.org/10.1111/j.1744-6570.2001.tb00100.x.

Reyes FA, Shuo D, and Yu H (2023) Shoulder-support exoskeletons for overhead work: Current state, challenges and future directions. *IEEE Transactions on Medical Robotics and Bionics* 5(3), 516-527. https://ieeexplore.ieee.org/document/10176317

Schiebl J, Maufroy C, Ziegenspeck N, Giers C, Elmakhzangy B, Schneider U, and Bauernhansl T (2025) Performance characterization of a novel semi-active exoskeleton for overhead work. *Wearable Technologies* 6, e36. https://doi.org/10.1017/wtc.2025.10019

Smith, EM (2024) Defining assistive technology: Adopting a common framework. *Assistive Technology*, 36(6), 397-397. https://doi.org/10.1080/10400435.2024.2416364

Sposito M, Fanti V, Poliero T, Caldwell DG, and Di Natali C (2024) Field assessment of active BSE: Trends over test days of subjective indicators and self-reported fatigue for railway
20

construction workers. *Heliyon*, 10(12), e33055.
https://doi.org/10.1016/j.heliyon.2024.e33055

Sreenivasan G, Zhu C and Yi J (2024) Exoskeleton-assisted balance and task evaluation during quiet stance and kneeling in construction. *arXiv preprint arXiv*:2408.07795. https://doi.org/10.48550/arXiv.2408.07795.

Søraa RA and Fosch-Villaronga E (2020) Exoskeletons for all: The interplay between exoskeletons, inclusion, gender, and intersectionality. *Paladyn, Journal of Behavioral Robotics* 11(1), 217-227. https://doi.org/10.1515/pjbr-2020-0036.

Zhu C and Yi J (2023) Knee exoskeleton-enabled balance control of human walking gait with unexpected foot slip. *IEEE Robotics and Automation Letters* 8(11), 7751-7758. https://doi.org/10.1109/LRA.2023.3322082.

Zhu Z, Dutta A and Dai F (2021) Exoskeletons for manual material handling – A review and implication for construction applications. *Automation in Construction* 122, 103493. https://doi.org/10.1016/j.autcon.2020.103493.

Zula K (2014) The future of nontraditional occupations for women: A comprehensive review of the literature and implications for workplace learning and performance. *Journal of Diversity Management* 9(1), 7-18.
21

Figure 1. Percentage of Construction Jobs Held by Women

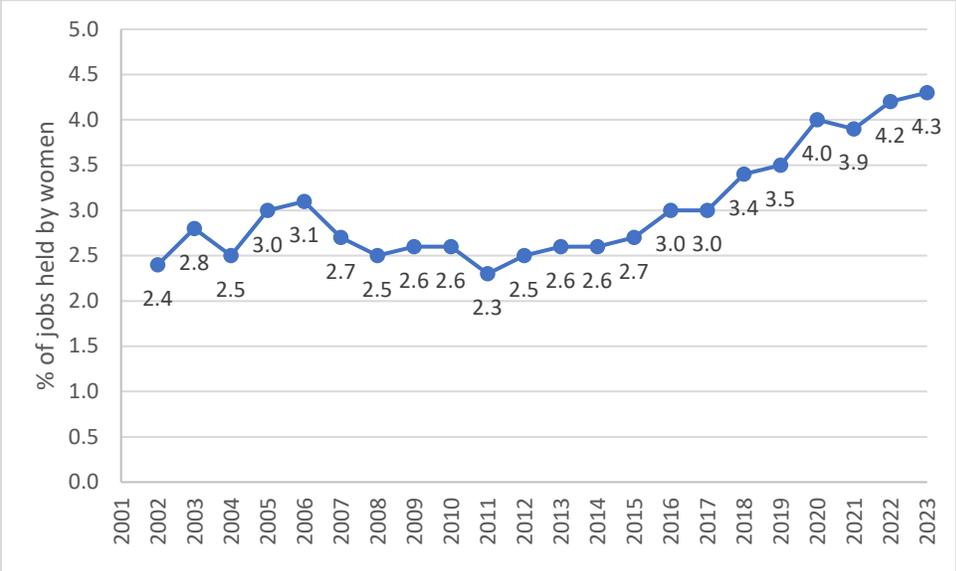

Source: Constructed by authors using U.S. Bureau of Labor Statistics, Current Population Survey, Table 11, for each year (BLS 2025a).

Table 1. Sample Means by Industry, SIPP Data

|  | Not Employed | Construction | Non-Construction |
|---|---|---|---|
| Woman | 0.601 | 0.083 | 0.499 |
|  | (0.490) | (0.276) | (0.500) |
| Mobility/strength impairments |  |  |  |
|   Climb 10 stairs | 0.207 | 0.016 | 0.031 |
|  | (0.405) | (0.127) | (0.174) |
|   Walk 3 blocks | 0.227 | 0.026 | 0.031 |
|  | (0.419) | (0.158) | (0.174) |
|   Stand 1 hour | 0.264 | 0.050 | 0.047 |
|  | (0.441) | (0.218) | (0.211) |
|   Sit 1 hour | 0.167 | 0.030 | 0.029 |
|  | (0.373) | (0.170) | (0.168) |
|   Stoop/crouch/kneel | 0.287 | 0.074 | 0.077 |
|  | (0.453) | (0.261) | (0.267) |
|   Reach overhead | 0.152 | 0.020 | 0.024 |
|  | (0.359) | (0.141) | (0.154) |
|   Lift/carry 10 pounds | 0.195 | 0.018 | 0.023 |
|  | (0.396) | (0.135) | (0.149) |
|   Grasp small objects | 0.109 | 0.017 | 0.019 |
|  | (0.311) | (0.128) | (0.135) |
|   Push/pull large objects | 0.246 | 0.036 | 0.045 |
|  | (0.430) | (0.186) | (0.206) |
| Other impairments |  |  |  |
|   Vision | 0.073 | 0.030 | 0.025 |
|  | (0.261) | (0.170) | (0.156) |
|   Hearing | 0.059 | 0.025 | 0.029 |
|  | (0.236) | (0.156) | (0.167) |
|   Speech | 0.043 | 0.010 | 0.005 |
|  | (0.204) | (0.098) | (0.074) |
|   Mental/cognitive | 0.181 | 0.066 | 0.052 |
|  | (0.385) | (0.249) | (0.222) |
| Education |  |  |  |
| High school or less | 0.518 | 0.605 | 0.301 |
|  | (0.500) | (0.489) | (0.459) |
| Some college/associate's degree | 0.292 | 0.259 | 0.291 |
|  | (0.455) | (0.438) | (0.454) |
| College degree | 0.131 | 0.091 | 0.252 |
|  | (0.337) | (0.288) | (0.434) |
| Post-college degree | 0.060 | 0.045 | 0.156 |
|  | (0.238) | (0.207) | (0.362) |
| Race/ethnicity |  |  |  |
| White non-Hispanic | 0.546 | 0.621 | 0.654 |
|  | (0.498) | (0.485) | (0.476) |
| Black non-Hispanic | 0.164 | 0.054 | 0.114 |
| | (0.370) | (0.225) | (0.318) |
| Hispanic | 0.191 | 0.279 | 0.151 |
| | (0.393) | (0.449) | (0.358) |
| Asian | 0.066 | 0.031 | 0.061 |
| | (0.247) | (0.174) | (0.239) |
| Other/Multiracial | 0.034 | 0.015 | 0.020 |
| | (0.181) | (0.121) | (0.141) |
| Age group | | | |
| 18-24 | 0.219 | 0.073 | 0.132 |
| | (0.414) | (0.260) | (0.339) |
| 25-34 | 0.174 | 0.235 | 0.235 |
| | (0.379) | (0.424) | (0.424) |
| 35-44 | 0.157 | 0.248 | 0.221 |
| | (0.364) | (0.432) | (0.415) |
| 45-54 | 0.180 | 0.292 | 0.232 |
| | (0.384) | (0.455) | (0.422) |
| 55-64 | 0.270 | 0.153 | 0.179 |
| | (0.444) | (0.360) | (0.383) |
| Ln(monthly earnings) | -- | 2.893 | 2.920 |
| | -- | (0.672) | (0.740) |
| Sample size | 6,554 | 1,001 | 12,591 |

Notes: Sample means for employed individuals using 2014 SIPP data. Standard deviations in parentheses. Means are weighted to population averages using SIPP sample weights.

Table 2. Gender and Mobility/Strength Gaps in Construction and Non-Construction Jobs, SIPP Data

|  | Employment | | Earnings | |
|---|---|---|---|---|
|  | Model 1: Construction | Model 2: Non-Construction | Model 3: Construction | Model 4: Non-Construction |
| Woman | -0.249*** | -0.079*** | -0.356*** | -0.231*** |
|  | (0.009) | (0.007) | (0.118) | (0.014) |
| Mobility/strength impairments | | | | |
|   Climb 10 stairs | -0.022 | -0.035* | 0.110 | (0.069) |
|  | (0.014) | (0.020) | -0.178 | -0.049 |
|   Walk 3 blocks | -0.055*** | -0.128*** | 0.442*** | -0.095* |
|  | (0.014) | (0.021) | -0.153 | -0.058 |
|   Stand 1 hour | -0.048*** | -0.145*** | (0.163) | 0.052 |
|  | (0.017) | (0.020) | -0.157 | -0.041 |
|   Sit 1 hour | 0.002 | 0.009 | 0.095 | 0.025 |
|  | (0.012) | (0.018) | -0.264 | -0.048 |
|   Stoop/crouch/kneel | -0.040** | -0.032** | -0.277** | (0.029) |
|  | (0.016) | (0.015) | -0.121 | -0.031 |
|   Reach overhead | -0.003 | -0.032* | 0.095 | (0.002) |
|  | (0.011) | (0.018) | -0.139 | -0.051 |
|   Lift/carry 10 pounds | 0.011 | -0.104*** | 0.031 | 0.013 |
|  | (0.015) | (0.021) | -0.224 | -0.057 |
|   Grasp small objects | -0.005 | -0.014 | (0.063) | -0.101* |
|  | (0.012) | (0.019) | -0.239 | -0.054 |
|   Push/pull large objects | -0.036*** | -0.092*** | 0.077 | (0.016) |
|  | (0.013) | (0.018) | -0.142 | -0.043 |
| Other impairments | | | | |
|   Vision | -0.012 | -0.057*** | 0.109 | -0.113* |
|  | (0.014) | (0.019) | (0.151) | (0.067) |
|   Hearing | -0.043*** | -0.009 | -0.194* | 0.010 |
|  | (0.014) | (0.017) | (0.111) | -0.048 |
|   Speech | -0.027 | -0.127*** | -0.88 | -0.178* |
|  | (0.023) | (0.025) | (0.769) | -0.1 |
|   Mental/cognitive | -0.045*** | -0.107*** | -0.149 | -0.105*** |
|  | (0.012) | (0.014) | (0.159) | -0.038 |
| Education (reference: high school or less) | | | | |
|   Some college/associate deg. | -0.019* | 0.093*** | 0.193*** | 0.199*** |
|  | (0.010) | (0.009) | (0.059) | -0.018 |
|   College degree | -0.078*** | 0.150*** | 0.315*** | 0.551*** |
|  | (0.013) | (0.010) | -0.111 | (0.019) |
|   Postgraduate degree | -0.069*** | 0.180*** | 0.535*** | 0.822*** |
|  | (0.021) | (0.011) | -0.116 | -0.022 |
| Race/ethnicity (reference: white non-Hispanic) | | | | |
| Black non-Hispanic | -0.101*** | -0.058*** | (0.222) | -0.121*** |
|  | (0.012) | (0.012) | -0.144 | -0.023 |
| Hispanic | 0.004 | -0.059*** | -0.368*** | -0.164*** |
| | (0.013) | (0.011) | (0.061) | (0.022) |
| Asian | -0.063*** | -0.126*** | 0.134 | -0.018 |
| | (0.017) | (0.017) | (0.163) | (0.034) |
| Other/Multiracial | -0.065*** | -0.086*** | -0.347** | (0.043) |
| | (0.019) | (0.022) | (0.166) | -0.041 |
| Age group (reference: 18-24) | | | | |
| 25-34 | 0.173*** | 0.151*** | 0.494*** | 0.385*** |
| | (0.015) | (0.014) | (0.126) | -0.03 |
| 35-44 | 0.225*** | 0.184*** | 0.737*** | 0.644*** |
| | (0.015) | (0.014) | (0.124) | -0.029 |
| 45-54 | 0.242*** | 0.210*** | 0.774*** | 0.685*** |
| | (0.015) | (0.014) | -0.125 | (0.028) |
| 55-64 | 0.140*** | 0.105*** | 0.827*** | 0.684*** |
| | (0.012) | (0.014) | -0.126 | (0.029) |
| Constant | 0.216*** | 0.604*** | 2.243*** | 2.188*** |
| | (0.012) | (0.013) | -0.124 | -0.028 |
| | | | | |
| Observations | 7,548 | 19,119 | 1,001 | 12,591 |

Notes: Estimates derived from 2014 SIPP data. Employment results are from linear probability regressions, and earnings results are from Heckman selection regressions of log hourly earnings. Standard errors in parentheses. *** statistically significant at 1%, ** at 5%, and * at 10% in 2-tail t tests.

Table 3. Physical ability levels, percent female, and annual earnings by occupation in construction

| Six-digit Occupations | Percent Female | Physical Level | Annual Income |
|---|---|---|---|
| Hazardous materials removal workers | 22.52 | 2.21 | 47,255 |
| Construction and building inspectors | 11.03 | 1.47 | 66,697 |
| Painters and paperhangers | 9.09 | 2.35 | 37,701 |
| Explosives workers, ordnance handling experts, and blasters | 7.07 | 2.20 | 55,750 |
| Helpers, construction trades | 6.35 | 2.65 | 29,509 |
| Sheet metal workers | 4.84 | 2.40 | 53,131 |
| First-line supervisors of construction and extraction workers | 4.18 | 1.84 | 76,052 |
| Other construction and related workers | 4.16 | 2.25 | 44,635 |
| Construction laborers | 4.10 | 2.78 | 39,797 |
| Solar photovoltaic installers | 3.94 | 2.25 | 43,244 |
| Insulation workers | 3.71 | 2.44 | 51,741 |
| Drywall installers, ceiling tile installers, and tapers | 3.48 | 2.46 | 40,986 |
| Carpet, floor, and tile installers and finishers | 3.28 | 2.41 | 42,980 |
| Highway maintenance workers | 3.27 | 2.36 | 46,454 |
| Underground mining machine operators | 2.77 | 2.51 | 74,573 |
| Other extraction workers | 2.71 | 2.40 | 62,043 |
| Roofers | 2.70 | 2.90 | 40,368 |
| Rail-track laying and maintenance equipment operators | 2.67 | 2.63 | 65,682 |
| Construction equipment operators | 2.65 | 2.04 | 59,131 |
| Carpenters | 2.49 | 2.69 | 46,153 |
| Electricians | 2.48 | 2.67 | 64,052 |
| Boilermakers | 2.33 | 2.32 | 69,704 |
| Fence erectors | 2.21 | 2.75 | 37,402 |
| Plumbers, pipefitters, and steamfitters | 2.02 | 2.60 | 59,937 |
| Glaziers | 1.82 | 2.18 | 51,626 |
| Structural iron and steel workers | 1.72 | 3.13 | 60,882 |
| Surface mining machine operators and earth drillers | 1.70 | 1.93 | 62,836 |
| Pipelayers | 1.63 | 2.35 | 52,292 |
| Plasterers and stucco masons | 1.44 | 2.69 | 42,009 |
| Derrick, rotary drill, and service unit operators | 1.41 | 2.40 | 73,838 |
| Elevator installers and repairers | 1.37 | 2.03 | 99,999 |
| Brick masons and reinforcing iron and rebar workers | 1.33 | 2.51 | 45,662 |
| Cement masons, concrete finishers, and terrazzo workers | 0.98 | 2.81 | 47,508 |

Note: Calculated by authors using data from the Occupational Information Network (O*NET) merged with 2018-2022 American Community Survey data.